\documentstyle[psfig,draft]{mn}

\def\gs{\mathrel{\raise0.35ex\hbox{$\scriptstyle >$}\kern-0.6em
\lower0.40ex\hbox{{$\scriptstyle \sim$}}}}
\def\ls{\mathrel{\raise0.35ex\hbox{$\scriptstyle <$}\kern-0.6em
\lower0.40ex\hbox{{$\scriptstyle \sim$}}}}

\title[Sub-mm observations of cooling-flow clusters]
      {The detection of dust in the central galaxies of distant cooling-flow
clusters}

\author[A.\,C.\ Edge et al.]
       {A.\,C.\ Edge,$^{\! 1}$ R.\,J.\ Ivison,$^{\! 2,3}$  Ian Smail,$^{\! 1}$
	A.\,W.\ Blain$^{4}$ \& J.-P.\ Kneib$^{5}$
        \vspace*{1mm}\\
        $^1$ Department of Physics, University of Durham, South Road,
        Durham DH1 3LE\\
        $^2$ Institute for Astronomy, Dept.\ of Physics \& Astronomy, 
        University of Edinburgh, Blackford Hill, Edinburgh EH9 3HJ\\
        $^3$ Department of Physics \& Astronomy, University College
        London, Gower Street, London WC1E 6BT\\
        $^4$ Cavendish Laboratory, Madingley Road, Cambridge CB3 0HE\\
        $^5$ Observatoire de Toulouse, 14 avenue E.\ Belin,
        31400 Toulouse, France}

\date{Accepted 2nd February 1999; Received 20th January 1999; in original 
form 7th September 1998}

\pagerange{000--000}

\begin{document}

\maketitle

\begin{abstract}
We present 850-$\mu$m observations with the SCUBA submillimetre
(sub-mm) camera of the central galaxies in seven concentrated clusters
of galaxies at redshifts between 0.19 and 0.41. We detect sub-mm
emission from the central galaxies in the rich clusters A\,1835 and
A\,2390, and present upper limits for the central galaxies in the
remaining five clusters. The two galaxies which we detect both exhibit
unusually blue UV-optical colours and lie in clusters which contain
massive cooling flows, $\geq 1000$\,M$_{\odot}$\,yr$^{-1}$. Moreover,
both galaxies host relatively strong radio sources. Focusing on
these two systems, we present new and archival radio---optical
observations to provide a detailed view of their spectral energy
distributions.  Our analysis indicates that sub-mm emission from the
central galaxy of A\,1835 can be best understood as arising from dust,
heated either by vigorous star formation or an obscured active galactic
nucleus. For the central galaxy of A\,2390, the sub-mm flux is
marginally consistent with an extrapolation of the
centimetre--millimetre emission from the luminous radio source that
lies in its core; although we cannot rule out an excess flux density
from dust emission comparable to that seen in A\,1835.  We present
details of our multi-wavelength observations and discuss the
implications of these data for the interpretation of star formation in
cooling-flow galaxies.
\end{abstract}


\section{Introduction}

%
%
\begin{table*}
\caption{The properties of the clusters in the sub-mm survey.}
\begin{tabular}{lccccccccl}
\noalign{\medskip \hrule \medskip}
Cluster & $z$ & $\alpha$ & $\delta$ & $t^1$ & $S_{850\mu{\rm m}}$ & $L_X^2$ & $T_X^2$ & $\stackrel{{\bf .}}{\rm M}^{3}$ & Comments \\
& & (J2000) & (J2000) & (ks) & (mJy) & (erg\,s$^{-1}$) & (keV) & (${\rm M}_\odot$ yr$^{-1}$) & \\
\noalign{\medskip \hrule \medskip}
Cl\,0024$+$16 & 0.39 & \hbox{$00^{\rm h} 26^{\rm m} 35.80^{\rm s}$} & \hbox{$+17^{\circ} 09' 41.0''$} & 15.6 & $<8.7$ & $2.0\times 10^{44}$ & ... & 96$^{+90}_{-53}$ & Multiple gE \\
\noalign{\smallskip}
A\,370 & 0.37 & \hbox{$02^{\rm h} 39^{\rm m} 53.18^{\rm s}$} & \hbox{$-01^{\circ} 35' 58.0''$} & 33.8 & $<5.4$ & $1.6\times 10^{45}$ & 7.2$^{+1.0}_{-0.8}$ & $<97$ & Southern D  \\
\noalign{\smallskip}
MS\,0440$+$02 & 0.19 & \hbox{$04^{\rm h} 43^{\rm m} 10.08^{\rm s}$} & \hbox{$+02^{\circ} 10' 18.0''$} & 35.8 & $<3.9$ & $3.9\times 10^{44}$ & 5.3$^{+1.3}_{-0.9}$ & 63$^{+103}_{-40}$ & Multiple gE  \\
\noalign{\smallskip}
A\,851 & 0.40 & \hbox{$09^{\rm h} 42^{\rm m} 56.68^{\rm s}$} & \hbox{$+46^{\circ} 59' 09.0''$} & 30.1 & $<5.0$ & $6.8\times 10^{44}$ & 6.7$^{+2.7}_{-1.7}$ & $<46$ & Two D gals \\
\noalign{\smallskip}
A\,1835 & 0.25 & \hbox{$14^{\rm h} 01^{\rm m} 02.11^{\rm s}$} & \hbox{$+02^{\circ} 52' 43.1''$} & 23.0 & $4.0\pm 1.2$ & $4.5\times 10^{45}$ & 9.8$^{+2.3}_{-1.3}$ & 1760$^{+520}_{-590}$ & cD galaxy \\
\noalign{\smallskip}
A\,2390 & 0.23 & \hbox{$21^{\rm h} 53^{\rm m} 36.76^{\rm s}$} & \hbox{$+17^{\circ} 41' 44.2''$} & 33.7 & ~$9.9\pm 1.7^4$ & $4.1\times 10^{45}$ & 14.5$^{+15.5}_{-5.2}$ & 1530$^{+580}_{-1110}$ & cD galaxy \\
\noalign{\smallskip}
Cl\,2244$-$02 & 0.33 & \hbox{$22^{\rm h} 47^{\rm m} 12.37^{\rm s}$} & \hbox{$-02^{\circ} 05' 44.0''$} & 25.6 & $<6.0$ & $1.3\times 10^{44}$ & ... & $<40$ & Multiple gE \\
\noalign{\smallskip \hrule}
\end{tabular}
\medskip
{\small
\begin{tabular}{ll}
1)  & Exposure time of the 850-$\mu$m SCUBA maps. \\
2)  & 2--10-keV X-ray luminosity and temperature (Mushotzky \& Scharf
1997; Allen 1998).  \\
3)  & Mass deposition rate in the cooling flow from X-ray spectroscopy (Allen 1998) for A1835 and A2390. Mass deposition rates for \\
    & all other clusters derived from archival ROSAT imaging data using a deprojection code (see White, Jones \& Forman 1997).  \\
4)  & Confirmed by a photometry-mode SCUBA observation at 850\,$\mu$m:
$9.8 \pm 2.0$\,mJy; weighted mean: $9.9 \pm 1.3$\,mJy.  \\
\end{tabular}
}
\end{table*}

The cooling time of hot X-ray-emitting gas in the central regions of
massive, relaxed clusters of galaxies can be substantially less than
the Hubble time.  The gas in these regions can thus cool and recombine,
initiating a cooling flow (Fabian \& Nulsen 1977; Cowie \& Binney
1977).  The ultimate fate of this cooling gas has been the subject of
an extensive and strongly contested debate (see Fabian 1994). 
The cold gas
is not detected in molecular form and so is inferred to reside in a
phase with $T_{\rm gas} << 100$\,{\sc k}. Calculations of the gas
properties are consistent with current observed limits (Ferland, Fabian
\& Johnstone 1994).  The metals in this cold and probably dense gas
are likely to condense to form dust grains, although the lifetime
of these grains could be severely limited due to spluttering from 
the X-rays emitted by the surrounding hot gas in the cluster core.

Observations of the central cluster galaxies in cooling-flow clusters
indicate that large numbers of 
stars are not formed from the reservoir of cold gas created
by the cooling-flow, although it has been suggested that the initial
mass function may be biased in favour of low-mass stars so that the
integrated stellar spectrum is difficult to detect (Fabian, Nulsen \&
Canizares 1982). However, many cooling-flow clusters contain optical
emission-line nebulae around their central galaxies, on scales of up to
100\,kpc, and their emission-line ratios (assuming Case~B recombination) 
suggest that considerable amounts of dust may be
present in the central regions of these clusters obscuring
any stars formed there (Hansen, J\"orgensen \&
N\"orgaard-Nielsen 1995; Allen 1995).  Dust has also been inferred from
direct detection of dust lanes in recent {\it Hubble Space Telescope}
({\it HST}) imaging of central cluster galaxies (e.g.\ A1795, McNamara
et al.\ 1996; Pinkney et al.\ 1996).  However, it is difficult to
quantify the total mass or temperature of this dust from the
observations, although the dust mass implied could be substantial
enough to obscure any young, blue stars being formed from the
cooling-flow gas.  The origin of the dust seen in these galaxies is
still an open issue, it could either originate from the cooled
gas clouds themselves or from the on-going star formation. Unfortunately,
it is difficult to observationally distinguish between these two
scenarios.

A direct detection of dust in cooling-flow galaxies and a more robust
estimate of its total mass and temperature would provide a substantial
insight into the physical processes occuring in the cooling-flow gas
(e.g.\ Fabian, Johnstone \& Daines 1994). For the temperatures inferred
for dust in cooling flows ($<40$\,{\sc k}, see O'Dea et al.\ 1994), 
the dust will radiate
predominantly in the sub-mm waveband; effort has therefore been
invested in obtaining a direct confirmation of the presence of dust in
cooling-flow galaxies using the UKT14 bolometer on the James Clerk
Maxwell Telescope (JCMT) at 850 and 1100\,$\mu$m.  Observations of 11
central galaxies in cooling-flow clusters at $z<0.1$ were undertaken by
Annis \& Jewitt (1993).  All detections were compatible with an
extrapolation of the centimetre-wave radio emission associated with the
central cluster galaxies, but these observations were able to set
useful limits on the total mass of dust in the central 10\,kpc of these
systems: $< 10^8 {\rm M}_\odot$. The only nearby cooling flow with a
sub-mm detection is that around NGC~1275 in Perseus (Gear et al.\ 1985;
Lester et al.\ 1995), although the interpretation in this source is
complicated by the strongly varying nuclear component.  Moreover, the
presence of the dust may be related to apparently on-going galaxy merger in
this system, which has been the subject of a long-running debate (Van
den Bergh 1977; Hu et al.\ 1983; Pedlar et al.\ 1990; Holtzman et
al.\ 1992; N\"orgaard-Nielsen et al.\ 1993).  

In the far-infrared
the {\it Infrared Astronomical Satellite} ({\it IRAS}) detected
a number of central cluster galaxies (Bregman, McNamara \& O'Connell 1990;
Grabelsky \& Ulmer 1990; Wise et al.\ 1993; Cox, Bregman \& Schombert 1995).
Although the number of detections was small and foreground and background
contamination effects make it difficult to draw conclusions as
to the presence of dust in individual central cluster galaxies,
Cox et al. (1995) conclude that up to 10 per cent of clusters contain
a central cluster galaxy with 10$^7$ ${\rm M}_\odot$ of dust.
At higher redshifts, {\it IRAS} detected the central
galaxy in a cooling-flow cluster at $z=0.44$, {\it IRAS}\,09104$+$4109
(Kleinmann et al.\ 1988; Fabian \& Crawford 1995), although this is
likely to host an obscured quasar nucleus (Hines \& Wills 1993).  Thus
a conservative summary of previous sub-mm observations of cooling-flow
galaxies is that all are ambigious as a result of modest sensitivity
and/or strong contamination from non-thermal emission.

With the introduction of sensitive bolometer cameras operating in the
sub-mm, in particular the Submillimetre Common-User Bolometer Array
(SCUBA) on the JCMT (Holland et al.\ 1999), our ability to detect
emission from cold dust has been greatly enhanced. In this paper we discuss
sub-mm observations of seven distant, concentrated galaxy clusters
at $z=0.19$--0.41.  These were observed as part of a deep sub-mm
survey of the distant Universe (Smail, Ivison \& Blain 1997; Smail et
al.\ 1998a) which we  briefly review in \S2. The deep maps obtained at
850\,$\mu$m have detected emission associated with the
central galaxies of two of the clusters, A\,1835 ($z=0.23$) and
A\,2390 ($z=0.25$), and to place upper limits on the emission
from the remaining five central galaxies. In \S3 and \S4, we combine
our sub-mm data for the central galaxies in A\,1835 and A\,2390 with
new and archival observations in the radio, infrared (IR) and optical
wavebands, providing a view of their spectral energy distributions
across a wide wavelength range.  We discuss our results and then
give our conclusions in \S5.
Throughout we assume $\Omega_0=1$ and $H_0 = 50$\,km\,s$^{-1}$\,Mpc$^{-1}$.

\section{The SCUBA survey of lensing clusters}

The deep 850-$\mu$m maps analysed in this work were obtained in a
survey of the distant Universe as seen through foreground galaxy
clusters. Data were obtaining during several observing sessions in
1997 and 1998. The survey concentrated on massive, lensing clusters to
take advantage of the gravitational amplification of all background
galaxies.

The survey resulted in the discovery of a large population of distant
star-forming galaxies, the properties of which are discussed in Smail
et al.\ (1997, 1998a, 1999), Ivison et al.\ (1998, 1999), Blain et al.\
(1999a, 1999b) and Barger et al.\ (1999).  Those papers also provide details of the acquisition,
reduction and analysis of the 850-$\mu$m maps and the identification
of the optical counterparts of the sub-mm sources. In Table~1, we list the
clusters, the exposure times and the flux densities of the central
galaxies measured in 30-arcsec-diameter apertures ($\sim 120$--200\,kpc
at the cluster redshifts), or 3-$\sigma$ upper limits.  We have
adopted a fixed metric aperture for the flux measurement to accomodate
the possibility of extended dust emission in the cluster core.
Smail et al.\ (1998a) review the likelihood of each sub-mm source
having a counterpart in deep optical images. For the two detections
presented here, the brightest cluster galaxy is the most probable
counterpart and any offset between the sub-mm source and galaxy
positions is consistent with the combined astrometric accuracy of the
SCUBA maps and the signal to noise of the detected source.

Only two of the clusters in the survey show detectable 850-$\mu$m
emission from their central galaxies: A\,1835 and A\,2390. These
galaxies lie in two of the most massive known cooling flows, both are
sources of strong radio emission, and both exhibit strong optical
emission lines (Allen et al.\ 1992; Le~Borgne et al.\ 1991).  The
presence of substantial amounts of dust in A\,1835 was suggested by the
analysis of optical emission-line ratios by Allen (1995) who estimated
an intrinsic reddening of $E(B-V)=0.49^{+0.17}_{-0.15}$.  Correcting
for this Allen estimated a star-formation rate from the
H$\alpha$ luminosity as high as $400\,{\rm M}_\odot$\,yr$^{-1}$. These
levels of reddening (and therefore dust) and star formation would make
this galaxy comparable to the most extreme starbursts.

The remaining five clusters do not show detectable 850-$\mu$m  emission
from their central galaxies. Due to modest atmospheric transmission the
sensitivities of our 450-$\mu$m SCUBA maps do not provide interesting
limits for these five clusters.  None of these five clusters contains a
large cooling flow (all have $\stackrel{.}{M} < 100 {\rm M}_\odot
$\,yr$^{-1}$ see Table~1); only one -- MS\,0440$+$02 -- is known to exhibit optical
line emission (Donahue et al.\ 1992) or radio emission comparable to
A\,1835 or A\,2390 (41.5\,mJy at 1.4\,GHz from NVSS Condon et
al.\ (1998) which sets upper limits of 3\,mJy for all other sources).

We note that neither of the two detected central galaxies are in the
clusters originally analysed by Smail, Ivison \& Blain (1997) and
furthermore these sources have been removed from any subsequent
analysis of sub-mm source counts (Blain et al.\ 1999a).

%
%
\begin{figure*}
\centerline{\psfig{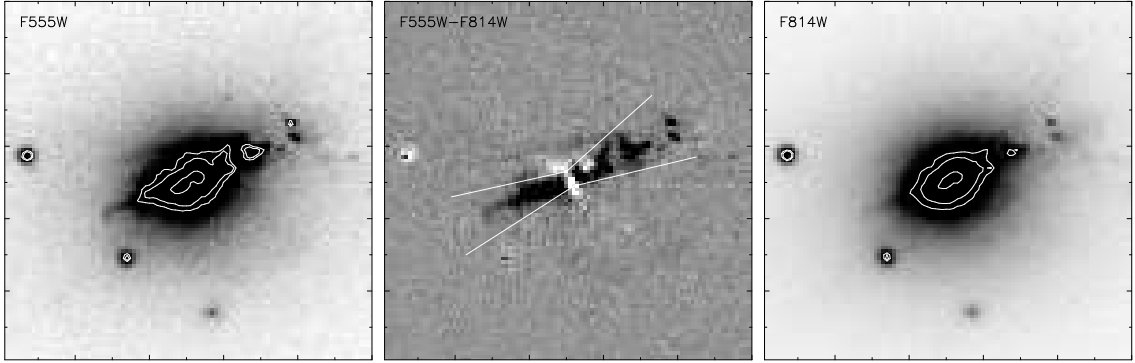} }
\caption{The archival {\it HST} WFPC2 images of the central galaxy in
A\,2390.  We show the F555W exposure on the right and the F814W on the
left.  There are obvious differences between the morphological
appearance of the galaxy in the two bands.  To emphasise these
differences, we have scaled and subtracted the F814W image from the
F555W exposure to remove the contribution from the red stellar halo of
the galaxy.  We show this colour-subtracted image in the central
panel, which shows the blue jet (black) and a red dust lane crossing
it (white).  The panels are each $10\times 10$\,arcsec square (47\,kpc
at the cluster redshift) and North is 19.3$^\circ$ counter-clockwise
from the vertical axis.  The thin white lines roughly
delimit the boundaries of the blue emission cone seen in the (F555W$-$F814W)
image.}
\end{figure*}

\section{Observations of the central galaxies in A\,1835 and A\,2390}

Here we review the available observations of the central galaxies in
A\,1835 and A\,2390. These include further sub-mm observations, new and
archival radio maps from the National Radio Astronomy Observatory
(NRAO) Very Large Array (VLA), as well as IR--optical photometry.  The
observations are summarised in Table~2.  Before discussing these in
detail we caution that the precise form of the spectral energy
distributions (SEDs) of these galaxies are sensitive to the choice of
aperture used in the different passbands.  Unfortunately without
full knowledge of the extent of the emission region as a function of
wavelength it is difficult to remove such effects from observations 
across a wide range of wavelengths and with different resolutions.  
In what follows we
have adopted the long wavelength measurements as total fluxes and
corrected the observations in the near-infrared and optical to `total'
values assuming the aperture correction derived for the cD galaxies in
the $I$-band.  This scheme assumes that the published {\it ISO} fluxes
in Lemonon et al.\ (1998) for A\,2390 corrected in the manner discussed
in that paper represent the total emission for this galaxy.

\subsection{Radio observations}

During 1997 October 22, we observed the central radio source in A\,2390
at frequencies between 1.4 and 43\,GHz using the D configuration of the
VLA. The data were reduced and analysed using standard {\sc aips}
procedures. In addition, we extracted three archival datasets from 1990
March 3, 1990 August 27 and 1995 May 10 to test for variability.  For
the 1990 August dataset, presented by L\'emonon et al.\ (1998), we
found no reliable phase solutions for any of the high-frequency data at
15 and 22\,GHz. The conclusions presented by L\'emonon et al.\ (1998),
based on these observations, are therefore likely to be inaccurate.
Using the remaining observations, we find no compelling evidence for
any variability of the nuclear component above the 10-per-cent level
between 1990 and 1998. 

The 1990 March and August data were obtained using
the VLA in configurations A and B, respectively, and so have
significantly higher resolution than the D-configuration maps from 1995
and 1997 (factors of 35 and 10 respectively). Most of the flux in the data
from 1990 is resolved at 4.89\,GHz into a $0.3''\times 0.15''$
component at a position angle of 44$^{\circ}$.  The higher spatial
resolution data results in a lower integrated flux density, indicating
that there is an extended component to the radio emission. This
extended emission is also evident from the detection of the source in
the Texas Survey at 365\,MHz (Douglas et al.\ 1996), where the
low-frequency tail of the second extended component exceeds the emission from
the inverted nuclear source. This
extended radio emission may be associated with a cluster halo, but more
sensitive low-frequency observations are required to investigate this
possibility.

For A\,1835, the radio data listed in Table~2 came from a number of
observing sessions with the VLA: the first, a 4.89-GHz map
in C configuration on 1994 November 14; the second, 1.4- and 15-GHz
maps in B configuration on 1997 February 13; the last, 1.4- and
8.4-GHz maps in A configuration on 1998 April 12. Data from Condon et
al.\ (1998) and Cooray et al.\ (1998) were included to give frequency
coverage comparable to A\,2390.  The wide variety of resolutions and
frequencies results in some scatter in the derived flux densities,  
since there is some extended radio emission in this source
on scales of 5--20$''$. However, most of 
the flux is in a compact ($<0.2''$) component, similar to but less powerful 
than that in A\,2390.  Within the limitations of comparing the 
different resolution data, the source does not appear to
be variable on timescales of a few years at a level required to significantly contribute
to the sub-mm.

\subsection{Further sub-mm observations}

To supplement the detections of the central galaxies in A\,1835 (450
and 850\,$\mu$m) and A\,2390 (850\,$\mu$m) in our sub-mm maps, we
obtained photometry-mode data using SCUBA (see Ivison et al.\ 1998 for
details of photometry-mode observations and data reduction). Data were
obtained at 1350\,$\mu$m for A\,1835 and at 450, 850, 1350 and
2000\,$\mu$m for A\,2390.

A 4-ks observation of A\,1835 at 1350\,$\mu$m was made under
relatively poor conditions (with a noise-equivalent flux density of
105\,mJy\,Hz$^{-1/2}$, nearly twice the nominal value) but we obtain a
useful upper limit of 3$\sigma < 4.9$\,mJy.

Observations of A\,2390 were made in excellent conditions. In total
exposure times of 0.9 and 1.8\,ks we obtained robust detections of the
emission from the central galaxy at 1350 and 2000\,$\mu$m
respectively.  At 450 and 850\,$\mu$m, a 1.8-ks exposure in
photometry mode confirmed the photometric accuracy of the relevant
Smail et al.\ (1999) maps. The weighted mean 850-$\mu$m flux density
is $9.9 \pm 1.3$\,mJy. The 450-$\mu$m upper limit (Table~2) is 
consistent with the 450-$\mu$m map of Smail et al.\ (1999).

%
%
\begin{table*}
\begin{center}
\caption{ \hfil The observed properties of the central galaxies of A\,2390
and A\,1835 from the radio to the ultraviolet waveband. \hfil }
\begin{tabular}{lcccl}
\noalign{\medskip \hrule \medskip}
Waveband & Telescope & A\,2390 & A\,1835 & {Comment} \cr
\noalign{\medskip \hrule \medskip}
$450\,\mu$m & JCMT & \hbox{$<28$\,mJy} & 20$\pm$5\,mJy  & Smail et al.\ (1999). \cr
$850\,\mu$m & JCMT & 9.9$\pm$1.2\,mJy & 4.0 $\pm$1.2\,mJy & Map-/photometry-mode weighted mean for A\,2390. \cr
$1350\,\mu$m & JCMT & 9.7$\pm$2.0\,mJy  &  \hbox{$<4.9$\,mJy} & Photometry mode.\cr
$2000\,\mu$m & JCMT &  13.8$\pm$3.8\,mJy & ... &  Photometry mode.\cr
\noalign{\smallskip}
$12\,\mu$m & {\it IRAS} & \hbox{$<99$\,mJy} & \hbox{$<129$\,mJy} &  \cr
$25\,\mu$m & {\it IRAS} & \hbox{$<159$\,mJy} & \hbox{$<204$\,mJy} &  \cr
$60\,\mu$m & {\it IRAS} & \hbox{$130 \pm 60$\,mJy} & \hbox{200$\pm$65\,mJy} &  \cr
$100\,\mu$m & {\it IRAS} & \hbox{$430 \pm 190$\,mJy} & \hbox{$<510$\,mJy} & \cr
\noalign{\smallskip}
$6.7\,\mu$m & {\it ISO}/CAM & \hbox{$300^{+50}_{-40}$\,$\mu$Jy} & ... & L\'emonon et al.\ (1998). \cr
$15\,\mu$m & {\it ISO}/CAM & \hbox{$500^{+80}_{-70}$\,$\mu$Jy} & ... & L\'emonon et al.\ (1998). \cr
\noalign{\smallskip}
$0.69$\,cm & VLA  & \hbox{$ 41.7 \pm 3.0$\,mJy} D &  ... & \cr
$1.05$\,cm & BIMA &   ...                          & \hbox{$ 3.31\pm0.94$mJy} & Cooray et al.\ (1998). \cr
$1.33$\,cm & VLA  & \hbox{$ 73.2 \pm 2.8$\,mJy} D &  ... & \cr
$2.01$\,cm & VLA  & \hbox{$101.1 \pm 1.5$\,mJy} D & \hbox{$  4.9\pm0.5$mJy} B & \cr
$3.53$\,cm & VLA  & \hbox{$139.4 \pm 2.7$\,mJy} D & \hbox{$  6.8\pm0.3$mJy} A & \cr
$6.17$\,cm & VLA  & \hbox{$210.8 \pm 0.9$\,mJy} D & \hbox{$ 13.0\pm0.3$mJy} C & \cr
$21.0$\,cm & VLA  & \hbox{$226.0 \pm 1.4$\,mJy} D & \hbox{$ 28.6\pm0.5$mJy} A & \cr
$20$\,cm   & NVSS & \hbox{$236.0 \pm 8.3$\,mJy} D  & \hbox{$41.4\pm 1.9$\,mJy} D & Condon  et al.\ (1998).\cr
$82$\,cm   & Texas & \hbox{$569 \pm 42$\,mJy} & ... & Douglas et al.\ (1996).\cr
\noalign{\bigskip}
$I_{\rm tot}$ & P200 & \hbox{$15.72 \pm 0.02$} & \hbox{$15.42 \pm 0.01$} & Total magnitude (Smail et al.\ 1998b). \cr
$U_{\rm ap}$  & P200 & \hbox{$20.03 \pm 0.03$} & \hbox{$19.06 \pm 0.02$} & Aperture magnitude (Smail et al.\ 1998b). \cr
$B_{\rm ap}$ & P200 & \hbox{$19.80 \pm 0.02$} &  \hbox{$19.27 \pm 0.01$} & Aperture magnitude (Smail et al.\ 1998b). \cr
$V_{\rm ap}$ & {\it HST} & \hbox{$18.74 \pm 0.02$} &  ... & Aperture magnitude (this paper). \cr
$I_{\rm ap}$ & P200 & \hbox{$17.40 \pm 0.02$} & \hbox{$17.30 \pm 0.01$} & Aperture magnitude (Smail et al.\ 1998b). \cr
$J_{\rm ap}$ & UKIRT & ... & $15.86 \pm 0.04$ & Aperture magnitude (this paper). \cr
$K_{\rm ap}$ & UKIRT & $14.88 \pm 0.03$ & $14.32 \pm 0.03$ & Aperture magnitude (this paper).  \cr
\noalign{\smallskip \hrule}
\end{tabular}
\end{center}
\end{table*}

\subsection{Far-infrared data}

To place limits on the emission from both central galaxies at
wavelengths between 12 and 100\,$\mu$m, we used the {\sc xscanpi}
facility at the Infrared Processing \& Analysis Center to analyse
co-added survey data from the {\it IRAS}. We found a 60-$\mu$m source
spatially coincident with the central galaxy of A\,1835. For A\,2390,
there were no formal detections, but there is tentative evidence ($\gs 2\sigma$) of
emission at 60 and 100\,$\mu$m. For completeness, we also list in Table~2 the 7-
and 15-$\mu$m flux densities for A\,2390 from the {\it Infrared Space
Observatory} ({\it ISO}) observations reported by L\'emonon et al.\ (1998).

\subsection{Optical and near-IR data}

Smail et al.\ (1998b) recently published a photometric survey of ten
moderate-redshift, X-ray-luminous clusters. They found that four of the
central cluster galaxies had very blue ultraviolet--optical colours;
bluer than the bulk of their red, spheroidal cluster members, and bluer
than the central galaxies in the remaining clusters. The most extreme
examples were the central galaxies in the two clusters discussed here,
A\,1835 and A\,2390. At rest-frame wavelengths of $\sim 3500$\,\AA,
they have luminosities a factor of two in excess of those expected from
their rest-frame optical fluxes assuming a normal giant elliptical
SED. We list the ultraviolet and optical
photometry for both cluster galaxies in Table~2.

The $UBI$ photometry from Smail et al.\ (1998b) includes total
$I$-band magnitudes and seeing-matched aperture photometry in $UBI$
within 3.0-arcsec-diameter apertures. These measurements have been
corrected for reddening assuming $E(B-V)=0.14$ towards A\,2390 and
$E(B-V)=0.05$ towards A\,1835. Photometric errors include an estimated
uncertainty in the reddening corrections of 10 per cent.

Near-infrared photometry for the central galaxies of A\,2390 and A\,1835 is 
listed in Table\,2. These data were obtained at the 3.8-m UK
Infrared Telescope (UKIRT) using IRCAM3, a 256$^2$ InSb array, on the
nights of 1998 July 11--16. The conditions were photometric and the
seeing varied from 0.8 to 1.2$''$. For photometric calibration, we
employed UKIRT faint standards (Casali \& Hawarden 1992).

In addition to the ground-based images described earlier, we have also
made use of archival {\it HST} WFPC2
images of A\,2390 (Pell\'o et al.\ 1999). These data comprise a 8.4-ks
integration in F555W ($V$) and a 10.8-ks integration in F814W ($I$)
and provide deep imaging of the central cluster galaxy at an effective
resolution of $\sim 0.5$\,kpc (Fig.~1).  Evidence for both dust and
the presence of a powerful nuclear source can be found in these images
(c.f.\ L\'emonon et al.\ 1998). A distinct bi-conical structure bisected
by a strong dust lane around 0.5~arcsec, or 1~kpc, in length is shown
in Fig.~1. The cones are significantly bluer than the galaxy as a
whole and have an opening angle of 20--30$^{\circ}$. These cones are
similar to those seen in Seyfert galaxies where they result from
direct ionization by the nuclear source.   Although the radio
emission is relatively amorphous, on the smallest
scales discernable from the VLA data there is weak evidence for a
component orientated orthogonally to the optical cones seen in the {\it
HST} images (c.f.\ A\,1795, McNamara et al.\ 1996, where these
components are aligned).

\section{Analysis and Results}

We now employ the observations outlined in \S3 to investigate the
nature of the sub-mm emission detected in the central galaxies of
A\,1835 and A\,2390.

\subsection{A\,1835}

The SED of the central galaxy of A\,1835 from radio wavelengths to the 
ultraviolet is shown in Fig.\,2.  To determine the possible non-thermal
contribution to the emission at sub-mm wavelengths we extrapolate the
synchrotron radio emission from 1.4---28\,GHz (with $\alpha = -0.8$,
where $F_{\nu} \propto \nu^{\alpha}$) into the sub-mm waveband.  This
extrapolation falls more than an order of magnitude short of the flux
density detected at 850\,$\mu$m and several orders of magnitude below
that at 450\,$\mu$m.  The  detection of modest emission at
60\,$\mu$m (\S3.3) is three orders of magnitude in excess of the
extrapolated radio continuum.  This suggests that most of the sub-mm
emission from the central galaxy in A\,1835 is due to thermal emission
from dust and represents the first unambiguous evidence for dust {\it
in emission} in the central regions of a cooling-flow cluster.

The far-IR and sub-mm data are fit adequately by the luminous
{\it IRAS} galaxy template SED compiled by Guiderdoni et al.\ (1998),
which demonstrates that warm dust heated by hot stars in a starburst
is a plausible emission mechanism, though we can not rule out some
contribution to the dust heating, possibly even a dominant
contribution, from the galaxy's active nucleus (see, for example,
Ivison et al.\ 1998, 1999).

The mass of dust estimated from these observations is extremely
sensitive to its assumed properties, particularly the temperature,
$T_{\rm d}$. We are helped here by the 60-$\mu$m detection of A\,1835,
which allows us to constrain $T_{\rm d}$ to $40 \pm 5$\,{\sc k} if the
dust remains optically thin throughout the far-IR and obeys an
emissivity law with index $+1.5$. Even if the dust becomes optically
thick at 100\,$\mu$m, $T_{\rm d}$ cannot rise above $50$\,{\sc k}.
This temperature is significantly above the value of $\sim 10$\,{\sc
k} expected for the cold gas condensing out of the cooling flow
(Fabian, Johnstone \& Daines 1994), though we cannot rule out a
substantial cold dust mass (with smaller masses of dust at higher
temperatures dominating the observed far-IR/sub-mm emission).
An upper limit to the cold dust component could be set by the
extinction of background objects behind the cluster. In particular
multiply-imaged gravitational arcs provide the opportunity to compare
the colours of several images of the same background object at
different positions within the cluster core.  

For $T_{\rm d} = 40$\,{\sc k}, we derive a dust mass, using standard
assumptions about dust properties, of $1.1 \times 10^8$\,M$_{\odot}$. 
This falls to $0.8 \times 10^8$\,M$_{\odot}$ if we allow $T_{\rm d}$
to rise to $50$\,{\sc k}.

We show the composite non-thermal and dust spectrum fit to our
observations in Fig.~2.  Integrating the flux between 10\,$\mu$m and
2\,cm from the dust component we derive an integrated flux of $1.3
\times 10^{-14}$\,W\,m$^{-2}$.  This translates into a far-IR
luminosity of $L_{\rm FIR} \sim 10^{12}$\,L$_{\odot}$, placing
the central galaxy in A\,1835 on the boundary of the class of
ultra-luminous IR galaxies (ULIRGs) defined in Saunders \& Mirabel
(1996).

Assuming that all of this far-IR emission is powered by star
formation, we can convert the $L_{\rm FIR}$ into a star-formation rate
(SFR). Depending upon the exact conversion recipe chosen this yields
SFRs of 140--240\,M$_{\odot}$\,yr$^{-1}$ (Leitherer \& Heckman 1995)
or 200--600\,M$_{\odot}$\,yr$^{-1}$ (Thronson \& Telesco 1986).  The
estimate of 200--400\,M$_{\odot}$\,yr$^{-1}$ based upon the H$\alpha$
luminosity and young stellar content of this galaxy (Allen 1995)
compares favourably with these values, considering the differences in
the assumed IMF.  Clearly, if the far-IR luminosity is supplied
by stars then this galaxy is undergoing a substantial episode of star
formation.

%
%
\begin{figure}
\centerline{\psfig{file=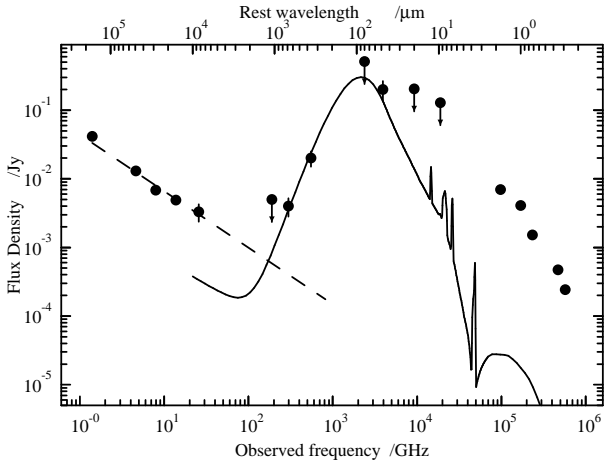,width=7.2cm}}
\caption{The SED of A\,1835 between the radio and ultraviolet
wavebands.  We plot the power-law extrapolated from the non-thermal
synchrotron radio emission as a dashed line
to illustrate the contribution expected
from this mechanism in the sub-mm waveband.  Note the substantial
excess emission in the sub-mm arising from dust. We also illustrate
the likely contribution from dust emission by plotting the composite
luminous {\em IRAS} SED compiled by Guiderdoni et al.\ (1998). The
excess over the model in the optical/near-ir is due to the old stellar
population in the massive host galaxy. }
\end{figure}

\subsection{A\,2390}

%
%
\begin{figure}
\centerline{\psfig{file=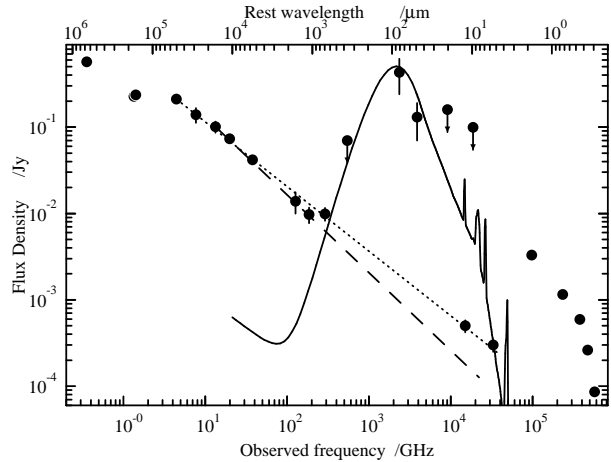,width=7.2cm}}
\caption{The SED of A\,2390 between the radio and ultraviolet
wavebands.  Again, we plot the extrapolated power-law based upon the
non-thermal synchrotron radio emission to illustrate the probable flux
in the sub-mm waveband. This is shown for both possible
extrapolations: the best fits between 2\,cm and 1350\,$\mu$m (dashed
line) and between 6\,cm and 7\,$\mu$m (dotted line).  We also
illustrate the likely contribution from dust emission by plotting the
composite luminous {\em IRAS} SED compiled by Guiderdoni et al.\
(1998). The excess at optical/near-IR wavelengths above the mid-IR
{\it ISO} data of L\'emonon et al.\ (1998) indicates that the {\it
ISO} measurements do not represent the total mid-IR fluxes of this
galaxy. This may be due (at least partially) to the use of small
apertures to measure the mid-IR flux.}
\end{figure}

The SED of the central galaxy in A\,2390 (Fig.~3) is quite different
from that of A\,1835, with a radio flux almost an order of
magnitude stronger -- the radio source in A\,2390 is powerful compared
to other central cluster galaxies in Abell clusters (Ledlow \& Owen 1995).  
Moreover, the radio spectrum shows signs of self-absorption
effects with a turnover at around 1--3\,GHz, indicating that the
radio source is probably very compact.

Turning to the mm/sub-mm detections, we see that these fall close to
an extrapolation of the radio spectrum arising from the non-thermal
nuclear source.  If the flux densities at 1350 and 2000\,$\mu$m are
due entirely to non-thermal processes, then the spectral index from those
points and the 15-, 22- and 43-GHz radio points is $\alpha = -0.88 \pm
0.08$.  The extrapolation of this power law to 850\,$\mu$m yields $6.5
\pm 1.0$\,mJy, and so the observed flux at 850\,$\mu$m then indicates a
marginally significant excess of $3.4 \pm 1.6$\,mJy.  This is
comparable with the dust emission in A\,1835 and hence the dust mass 
and SFRs infered would also be similar.

An alternative extrapolation of the non-thermal emission
would be a power law of the form $\alpha = -0.73 \pm 0.03$, which can
account for all the observed data points from 7\,$\mu$m to 5\,cm, and
only over-predicts the 1350- and 2000-$\mu$m fluxes at the 2- and
1-$\sigma$ levels, respectively.  This second model of the non-thermal
emission would leave no room for a substantial contribution from dust
emission at 850\,$\mu$m. This conclusion is not significantly affected
by relative calibration errors between the radio and sub-mm as the
observed excess is found within sub-mm band alone where the same
planetary calibrator is used.

Without a robust detection of emission at shorter wavelengths (e.g.\
450\,$\mu$m) we cannot differentiate between the two possible models
of the non-thermal contribution to the sub-mm emission and prove
whether we see dust emission from this galaxy.  However, the tentative
{\it IRAS} detections at 60 and 100\,$\mu$m do support the presence of
some warm dust and the {\it HST} images clearly show that some dust is
present.  Adopting an $\alpha = -0.88$ power-law and assuming that the
dust properties are similar to those inferred for the central galaxy
of A\,1835, we would expect a dust mass of around $0.6-0.8 \times
10^8$\,M$_{\odot}$ in A\,2390.

A claim has recently been made for substantial dust emission from the
central galaxy of A\,2390 using mid-IR observations from {\it
ISO} at 7 and 15\,$\mu$m (L\'emonon et al.\ 1998, L98).  We can conclude that
the excess of flux at 15\,$\mu$m is very probably due to dust
but L98 severely underpredict the potential non-thermal contribution
due to the problems with phase calibration in the VLA data used by
L98 -- see \S3.1. Therefore additional evidence for dust in A2390 exists
but the ambiguity introduced by the non-thermal continuum and the
uncertainties in dust opacity, temperature and aperature
effects prevent us from drawing
any quantitative conclusions from the {\it ISO} data alone.

To further clarify the nature of the sub-mm emission mechanism in A\,2390
observations in the far-IR are required.  Using SCUBA at
450\,$\mu$m, the expected flux is around 20\,mJy if dust is present,
and a detection at this level is feasible in good conditions.
Alternatively, a PHT observation from {\it ISO} at 90\,$\mu$m exists
(P.I.\ Cesarsky). This might optimistically reach an rms level of
10\,mJy\,beam$^{-1}$ and so provide confirmation of our tentative
850-$\mu$m dust excess.

The case of A\,2390 clearly illustrates the problems that can be
encountered if only a few data points are available in an SED
and at least two
emission mechanisms are at play. Even with the extensive dataset
presented here we can only present at best tentative evidence of dust
emission in this galaxy.  This is a cautionary lesson for future
observers to build as broader spectrum as possible before concluding
dust emission is present.

\section{Discussion and Conclusions}

The observation of sub-mm emission in the central galaxy of A\,1835
represents the first unambigious detection of dust {\it emission} in
the core of a cooling flow.  It provides an estimate of the total dust
mass in the core of this very massive cooling flow of around
$10^8\,{\rm M}_\odot$.  This dust is detected within a radius of
70\,kpc of the cluster centre where $\sim 10^{12}\,{\rm M}_\odot$ of
cold gas could have been deposited, if the cooling flow has existed in
its present state for the last 10\,Gyr (Allen et al.\ 1996).  A
minimum lifetime for the dust in this region is set by the timescale
for destruction due to spluttering by X-rays, which is only
0.1--1.0\,Gyr (Dwek \& Arendt 1992). However, the dust may reside in a
shielded environment resulting in a much longer lifetime.
Nevertheless, assuming that the dust is associated with the cooling
gas and that the dust will only survive for 0.1\,Gyr, the
corresponding gas-to-dust ratio could be around $10^3$, an order of
magnitude greater than the standard Galactic value.  Therefore,
despite obtaining a reliable measure of the dust mass, it is not
possible to set any firm limits on the total gas mass as the timescales 
for both the production and destruction of dust are short and the origin of 
the dust is unclear.

The observed dust temperature of $\sim 40$\,{\sc k} is significantly warmer
than would be expected from cold clouds (Ferland et al.\ 1994), and so it
is likely that the dust is warmed by, or is formed in, regions of star
formation. The presence of an additional, dominant, cold dust
component is possible though, and it is difficult to set an upper
limit since any warm dust will inevitable dominate the observed
emission (Frayer et al.\ 1999).

The vigorous star formation implied for A\,1835 is supported by the
tentative detection of Wolf-Rayet features in the optical spectrum
(Allen 1995). The short lifetimes of these WR stars, coupled with the
exceptional size of the cooling flow in A\,1835, implies that these
massive stars may be constantly replenished and that this system is a
sustained starburst.  The far-IR luminosity of the central galaxy in
A\,1835 classes it as a ULIRG, while in A\,2390 the exact
classification depends upon the interpretation of the non-thermal
component. The properties of both these galaxies may be more useful to
our understanding of dusty, strongly star-forming galaxies at earlier
epochs --- particularly those in high density environments --- than
more luminous, but isolated local ULIRGs (e.g.\ Arp\,220).  In
particular, it is often argued that luminous radio sources at high
redshift inhabit high-density environments (Hughes \& Dunlop 1998) and
may have associated cooling flows (Fabian et al.\ 1986).  The recent
SCUBA detection by Best et al.\ (1998) of 3C\,324, a radio galaxy at
$z=$1.2 within an X-ray luminous cluster (Smail \& Dickinson 1995;
Crawford \& Fabian 1996) illustrates this.  Thus, the nature of these
distant radio galaxies might be best probed through the detailed study
of the star formation in galaxies such as A\,1835 and A\,2390. 

This work illustrates the difficulty in detecting dust in central
galaxies of cooling flows.  Specifically, we have highlighted both the
pitfalls of interpreting observations of
central galaxies which contain strong radio
sources and of relying on limited radio coverage.  AJ93
demonstrated the problems for a number of lower redshift clusters where
each of their detections was attributable to the associated radio
source.  Even for weak radio sources, say fainter than 50\,mJy at
1.4\,GHz, at least one frequency above 5\,GHz should be mapped before
sub-mm emission can be confidently associated with dust.   Fortunately,
the majority of central cluster galaxies fall below this flux density
limit so future studies will be able to avoid the worst of these
problems.  

However, our detection of relatively weak sub-mm emission from only two galaxies in a
sample of seven implies that substantial effort will be required to
detect other central cluster galaxies and that it is unlikely that dust
emission will be detectable in typical central cluster galaxies
{\it at low redshift} with current instrumentation.  However, by
combining detailed spectroscopy and sub-mm photometry and focusing on a
well defined sample of the most massive cooling flow clusters we may
achieve a clearer understanding of the physical processes responsible
for star formation in cooling flow galaxies.

In the longer term with the advent of large sub-mm interferometer
arrays the number of detectable central galaxies should increase dramatically.
The detection of dust masses as low as 10$^6 {\rm M}_\odot$ will be
possible in relatively short exposures, allowing the study of less
extreme cooling flows.  Detections at shorter wavelengths,
50--350\,$\mu$m, will also provide more stringent constraints to be
imposed on the mass and temperature of the dust. These wavelengths are
inaccessable currently but will be easily within the capabilities of
{\it SIRTF}, {\it SOFIA} and {\it FIRST}.

To conclude, the detection of dust emission in the sub-mm from
the core of a cooling flow has revealed an additional and substantial
baryonic component in this dense region.  Future work will reveal how
representative this detection is, and how it relates to the deposition
of gas within cooling flows in general.

\subsection*{ACKNOWLEDGEMENTS}

ACE and IRS acknowledge support through a Royal Society University Research
Fellowships. RJI acknowledges the award of a PPARC Advanced
Fellowship. We thank Richard Ellis and Roger Blandford for discussions
and David White for use of the X-ray deprojection code.  UKIRT and
JCMT are operated by the Joint Astronomy Centre on behalf of the
PPARC, the Netherlands Organisation for Scientific Research and the
National Research Council of Canada.  Based on observations with the
NASA/ESA {\it Hubble Space Telescope} obtained at the Space Telescope
Science Institute, which is operated by the Association of
Universities for Research in Astronomy Inc., under NASA contract NAS
5-26555. NRAO is a facility of the National Science Foundation
operated under cooperative agreement by Associated Universities, Inc.

\end{document}